\documentclass[final,5p,times,twocolumn,nopreprintline]{elsarticle}
\usepackage{amsmath,slashed,booktabs,dsfont}
\usepackage{graphicx,graphics}
\usepackage{dcolumn}
\usepackage[hyperfootnotes=false]{hyperref}
\usepackage{xspace}
\usepackage{color}
\usepackage{balance}
\usepackage{multirow}
\usepackage{multicol}

\usepackage{fancyhdr}
\fancypagestyle{firstpage}{%
	
	\lhead{}
	\rhead{}
}
\newcommand{\GeV}{\,\text{GeV}}

\newcommand{\Order}{\mathcal{O}}
\newcommand{\Br}{\text{Br}}

\renewcommand{\Im}{\text{Im}\,}

\allowdisplaybreaks[1]

\begin{document}

\renewcommand{\theequation}{\arabic{equation}}

\begin{frontmatter}
 
\title{On the leptonic decays of the $\tau$}

\author[Bern]{Martin Hoferichter}
\author[Pisa]{Alberto Lusiani}
\author[Liverpool]{Yannick Ulrich}

\address[Bern]{Albert Einstein Center for Fundamental Physics, Institute for Theoretical Physics, University of Bern, Sidlerstrasse 5, 3012 Bern, Switzerland}

\address[Pisa]{INFN, Sezione di Pisa and Scuola Normale Superiore, Pisa, Italy}

\address[Liverpool]{Department of Mathematical Sciences, University of Liverpool, Liverpool, L69 3BX, United Kingdom}

\begin{abstract}
In view of potential future improvements in the branching fractions of leptonic $\tau$ decays $\tau\to\ell\nu_\tau\bar\nu_\ell$, $\ell=e,\mu$, it is timely to derive the corresponding Standard-Model predictions beyond next-to-leading order. We provide such benchmark values for the fully inclusive decay widths, profiting from higher-order corrections that have been studied in the $\overline{\text{MS}}$ scheme up to three-loop order in the case of muon decay. To obtain complete next-to-next-to-leading-order results, we perform a numerical calculation of the two-loop coefficients in the on-shell scheme, including an evaluation of the hadronic effects, and compare the leptonic results to the known $\overline{\text{MS}}$ coefficients where available. Our final results are presented in the $\overline{\text{MS}}$ scheme, to be able to also include an estimate of the three-loop contribution. In addition, we comment on radiative and rare decays, whose consideration is critical for the conversion between $\overline{\text{MS}}$ and on-shell results. 
\end{abstract}

\end{frontmatter}

\thispagestyle{firstpage}

\section{Introduction}

The leptonic branching fractions of the $\tau$ lepton have been measured to 
\begin{align}
  \Br[\tau\to \mu \nu_\tau\bar \nu_\mu]&=0.1736(4)&&\text{\cite{DELPHI:1999ldo,L3:2001szz,OPAL:2002xiz,ALEPH:2005qgp}},\notag\\
  \Br[\tau\to e \nu_\tau\bar \nu_e]&= 0.1784(4)&&\text{\cite{CLEO:1996oro,OPAL:1998lid,DELPHI:1999ldo,L3:2001szz,ALEPH:2005qgp}},
\end{align}
according to the averages of Ref.~\cite{HeavyFlavorAveragingGroupHFLAV:2024ctg},\footnote{In both cases, the last digit increases by one unit if the average including the unitarity constraint that all branching fractions sum to unity is imposed.} and the ratio is known to~\cite{HeavyFlavorAveragingGroupHFLAV:2024ctg}
\begin{equation}
  \frac{\Br[\tau\to \mu \nu_\tau\bar \nu_\mu]}{ \Br[\tau\to e \nu_\tau\bar \nu_e]}
  = 0.9730(22)\qquad \text{\cite{ARGUS:1991zhv,CLEO:1996oro,BaBar:2009lyd,Belle-II:2024vvr}}. 
\end{equation}
Given that the experimental precision could potentially improve by an order of magnitude at FCC-ee~\cite{FCC:2018byv}, it is timely to reassess the corresponding Standard-Model (SM) predictions, building on the detailed work that has been performed for muon decay. 
In this case, the precision requirement is set by the MuLan lifetime measurement $\tau_\mu=2.196 9803(22)\times 10^{-6}\,\text{s}$~\cite{MuLan:2010shf,MuLan:2012sih}, to be able to extract the Fermi constant $G_F\equiv G_F^{\mu e}=1.1663786(6)\times 10^{-5}\GeV^{-2}$~\cite{MuLan:2012sih,Eberhart:2026klz,ParticleDataGroup:2026mpi,Engel:2019nfw,MuLan:2007qkz,FAST:2007rsc} at a comparable level of precision.

For muon decay, the current status can be summarized as follows: radiative corrections in the $m_e=0$ limit are known at $\Order(\alpha)$~\cite{Kinoshita:1958ru}, $\Order(\alpha^2)$~\cite{vanRitbergen:1998yd,vanRitbergen:1999fi,Steinhauser:1999bx}, and $\Order(\alpha^3)$~\cite{Fael:2020tow,Czakon:2021ybq}, mass corrections were evaluated in Refs.~\cite{Nir:1989rm, Pak:2008qt,Pak:2008cp}, and hadronic effects, due to hadronic vacuum polarization (HVP), in Refs.~\cite{vanRitbergen:1998hn,Eberhart:2026klz}. While, in general, electroweak corrections are absorbed by convention into the definition of the Fermi constant, corrections from the expansion of the $W$ propagator remain~\cite{Ferroglia:2013dga,Fael:2013pja}, since they are not identical for different leptonic processes. In Sec.~\ref{sec:decay_rate_muon} we dissect the resulting prediction for muon decay and adapt it wherever possible to leptonic $\tau$ decays. For most contributions, results in the $\overline{\text{MS}}$ scheme can be inferred from the literature, but, to the best of our knowledge, mass corrections to the internal muon loop in $\tau\to e\nu_\tau\bar\nu_e$ have not been calculated so far. We close this gap by performing a numerical calculation in the on-shell scheme~\cite{Engel:2019nfw} (based on numerical data generated using the {\sc McMule} Monte Carlo code~\cite{Banerjee:2020rww}), mapping the results to $\overline{\text{MS}}$ as detailed in Sec.~\ref{sec:leptonic}. We provide an evaluation of the hadronic contributions in Sec.~\ref{sec:hadronic}, leading to the complete SM predictions given in Sec.~\ref{sec:SM_prediction}. Some aspects of radiative and rare decays are addressed in Sec.~\ref{sec:radiative_rare},
before concluding in Sec.~\ref{sec:conclusions}.

\section{Muon decay}
\label{sec:decay_rate_muon}

We first review the SM prediction for muon decay, before adapting the result to leptonic $\tau$ decays. 
 The inclusive decay rate for muon decay can be written as\footnote{We follow mostly the conventions of Ref.~\cite{ParticleDataGroup:2026mpi}, but do not factorize the tree-level mass dependence $F(\rho)$, as this leads to more compact expressions at subleading orders.}
\begin{equation}
\label{muon_decay}
 \Gamma[\mu\to e\nu_\mu\bar\nu_e]=\Gamma_0^{\mu}\delta_{\mu e} \bigg[F(\rho_{\mu e})+\sum_{i\geq 1}H_i^{\mu e}(\rho_{\mu e})\bigg(\frac{\hat\alpha(m_\mu^2)}{\pi}\bigg)^i\bigg],
 \end{equation}
 where
 \begin{align}
 F(\rho)&=1-8\rho+8\rho^3-\rho^4-12\rho^2\log\rho,\notag\\
 H_1^{\mu e}(\rho)&=\frac{25}{8}-\frac{\pi^2}{2}-\big(34+12\log\rho\big)\rho+16\pi^2\rho^{3/2}\notag\\
 &-\bigg(\frac{273}{2}+8\pi^2-18\log \rho+18\log^2\rho\bigg)\rho^2\notag\\
 &+16\pi^2\rho^{5/2}-\bigg(\frac{526}{9}-\frac{76}{3}\log\rho\bigg)\rho^3+\Order\big(\rho^{7/2}\big),\notag\\
 H_2^{\mu e}(\rho)&=\frac{156815}{5184}-\frac{518}{81}\pi^2-\frac{895}{36}\zeta(3)+\frac{67}{720}\pi^4+\frac{53}{6}\pi^2\log 2\notag\\
 &+H_2^{\mu e, \,\text{HVP}}-\frac{5}{4}\pi^2\sqrt{\rho}
+\bigg(-\frac{86189}{360}+\frac{1361}{108}\pi^2+86\zeta(3)\notag\\
&+\frac{121}{540}\pi^4-8\pi^2\log 2-\frac{263}{6}\log\rho-11\log^2\rho\bigg)\rho
 \notag\\
 &+\Order(\rho^{3/2}),
\end{align}
and
\begin{equation}
\label{Gamma_0}
 \Gamma_0^{\ell}=\frac{G_F^2m_\ell^5}{192\pi^3},\qquad 
 \delta_{\ell\ell'}=1+\frac{3}{5}\frac{m_\ell^2+3 m_{\ell'}^2}{M_W^2},\qquad \rho_{\ell\ell'}=\frac{m_{\ell'}^2}{m_\ell^2}.
\end{equation}
$H_{1,2}^{\mu e}(0)$ were computed in Refs.~\cite{Kinoshita:1958ru,vanRitbergen:1998yd,vanRitbergen:1999fi,Steinhauser:1999bx}, respectively, including the HVP contribution $H_2^{\mu e,\, \text{HVP}}=-0.042(2)$~\cite{vanRitbergen:1998hn},\footnote{This value agrees with our evaluation given in Eq.~\eqref{HVP_muon} below, which in turn is very close to the recent evaluation $H_2^{\mu e,\, \text{HVP}}=-0.0428(6)$~\cite{Eberhart:2026klz}.} and mass corrections in Refs.~\cite{Nir:1989rm, Pak:2008qt,Pak:2008cp}.  An estimate of the third-order term 
\begin{equation}
\label{three-loop}
H_3^{\mu e}=-15.3(2.3)
\end{equation}
was provided in Ref.~\cite{Fael:2020tow}, with parts of the calculation independently verified in Ref.~\cite{Czakon:2021ybq}. 
While most electroweak corrections are, by definition, absorbed into the Fermi constant $G_F$, process-dependent mass corrections from the expansion of the $W$ propagator~\cite{Ferroglia:2013dga,Fael:2013pja} need to be added as given in Eq.~\eqref{Gamma_0}. Since, after Fierz rearrangement, the charged lepton part of the muon decay operator corresponds to a conserved current, it can be defined in such a way that its anomalous dimension vanishes~\cite{Jenkins:2017jig,Jenkins:2017dyc,Naterop:2025cwg}, so that in such a scheme no further renormalization group corrections arise and $G_F$ in Eq.~\eqref{Gamma_0} can indeed be identified with $G_F^{\mu e}$.

The $\Order(\alpha^2)$ contribution decomposes into~\cite{vanRitbergen:1999fi,vanRitbergen:1999gs}
\begin{align}
H_2^{\mu e}(0)&=H_2^\gamma+H_2^e+H_2^\mu+H_2^{\mu e,\,\text{HVP}},\notag\\
 H_2^\gamma&=\frac{11047}{2592}-\frac{515}{81}\pi^2-\frac{223}{36}\zeta(3)+\frac{67}{720}\pi^4+\frac{53}{6}\pi^2\log 2,\notag\\
 H_2^e&=-\frac{1009}{288}+\frac{77}{216}\pi^2+\frac{8}{3}\zeta(3),\notag\\
 H_2^\mu&=\frac{16987}{576}-\frac{85}{216}\pi^2-\frac{64}{3}\zeta(3),
\end{align}
corresponding to purely photonic diagrams, diagrams containing an electron loop or an $e^+e^-$ pair in the final state, and diagrams containing muon loops, respectively.
$\tau$ loops decouple as $m_\mu^2/m_\tau^2$, and can therefore be neglected at the current level of precision~\cite{vanRitbergen:1998hn}. The $\Order(\alpha^3)$ coefficient is again defined fully inclusively, but actually calculated via an expansion in $\delta=1-m_e/m_\mu$~\cite{Fael:2020tow}, which converges surprisingly well even for $\delta=1$. The uncertainty estimate is derived by comparison to the convergence pattern when applying the same procedure to the $\Order(\alpha^2)$ coefficient.

For consistency with higher orders, it is crucial that all contributions be evaluated in the same scheme. However, while for evaluations in the massless limit the $\overline{\text{MS}}$ scheme is most useful, the HVP contribution (and the $\tau$ loop) are most easily evaluated in the on-shell scheme~\cite{vanRitbergen:1998hn}. As shown in Ref.~\cite{vanRitbergen:1999fi}, these calculations can be combined consistently if the $\overline{\text{MS}}$ coupling $\hat \alpha(\mu^2)$ in Eq.~\eqref{muon_decay} is replaced in favor of the on-shell Thomson limit $1/\alpha=137.0359991(1)$ by means of
\begin{align}
\label{alpha_conversion}
 \hat\alpha(m_\mu^2)&\to\alpha-\alpha^2\bigg(\frac{1}{3\pi}+\frac{\alpha}{4\pi^2}\bigg)\log\rho_{\mu e}+\frac{15\alpha^3}{8\pi^2}\\
 &\to \frac{\alpha}{1+\frac{\alpha}{3\pi}\log\rho_{\mu e}}-\frac{\alpha^3}{4\pi^2}\log\rho_{\mu e}+\frac{15\alpha^3}{8\pi^2}=\frac{1}{135.901},\notag
\end{align}
where the second replacement resums logarithms of the form $\alpha^n\log^{n-1}\rho_{\mu e}$, keeping as well the $\alpha^3\log\rho_{\mu e}$ term without producing spurious logarithms at higher order. The finite terms at $\Order(\alpha^3)$ are kept for both the electron and the muon loop, for consistency with the three-loop result in the $\overline{\text{MS}}$ scheme from Ref.~\cite{Fael:2020tow}, reproducing the numerical value of $\hat\alpha(m_\mu^2)$ from Ref.~\cite{ParticleDataGroup:2026mpi}. While for a full evaluation also HVP effects at $\Order(\alpha^3)$ would need to be considered, already the leptonic three-loop corrections are suppressed by a factor of $30$ compared to the experimental uncertainty.

\section{Leptonic contributions}
\label{sec:leptonic}

For $\tau$ decays, the following modifications of the leptonic contributions are required. At $\Order(\alpha)$ and $\Order(\alpha^2)$ one has 
 \begin{align}
  H_1^{\tau\ell}(\rho)&=H_1^{\mu e}(\rho),\notag\\
  H_2^{\tau\ell}(\rho)&=H_2^\gamma+2H_2^e+H_2^\mu
  +H_2^{\tau \ell,\,\text{HVP}}+\Delta H_2^{\tau\ell}(\rho),
 \end{align}
where the HVP contribution will be addressed in Sec.~\ref{sec:hadronic} and $\Delta H_2^{\tau\ell}(\rho)$ encodes the mass-dependent changes. Strictly speaking, this correction depends on both $\rho_{\tau e}$ and $\rho_{\tau\mu}$ for both decay channels, but in practice only  corrections involving $\rho_{\tau\mu}$ matter. For the muon channel, these corrections can be inferred from Refs.~\cite{Pak:2008qt,Pak:2008cp}, whose analytic expressions we use up to the order in $\rho$ provided. 
The resulting two-loop coefficients are given in the first panel of Table~\ref{tab:MSbar_onshell}.
In addition, repeating the evaluation of the three-loop coefficient from Ref.~\cite{Fael:2020tow} for two instead of one massless lepton species gives 
\begin{equation}
    H_3^{\tau\ell}=-40(1),
\end{equation}
where the convergence in $\delta$ further improves compared to Eq.~\eqref{three-loop}. 
These coefficients need to be combined with the consistent version of $\hat\alpha(m_\tau^2)$, generalizing Eq.~\eqref{alpha_conversion} to 
\begin{align}
\label{alpha_conversion_tau}
 \hat\alpha(m_\tau^2)
 &\to \frac{\alpha}{1+\frac{\alpha}{3\pi}\big[\log\rho_{\tau \mu}+\log\rho_{\tau e}\big]}\\
 &-\frac{\alpha^3}{4\pi^2}\big[\log\rho_{\tau \mu}+\log\rho_{\tau e}\big]+\frac{45\alpha^3}{16\pi^2}=\frac{1}{134.701}.\notag
\end{align}

\begin{table}[t]
\scalebox{0.95}{
\renewcommand{\arraystretch}{1.3}
\centering
\begin{tabular}{lrr}
\toprule
       & $\Gamma[\tau\to\mu\nu_\tau\bar\nu_\mu + n\gamma]$    & $\Gamma[\tau\to e\nu_\tau\bar\nu_e + n\gamma]$      \\\midrule
 LO                             & $0.972561(5)$   & $0.999999(4)$                  \\
 NLO                            & $0.968686(5)$     & $0.995796(5)$                 \\
NNLO                         & $0.968553(5)$  & $0.995536(5)$                      \\\midrule
 NLO/LO                             & $0.996016(7)$    & $0.995798(6)$                   \\
 NNLO/NLO                             & $0.999863(7)$    & $0.999739(6)$                   \\
\midrule
 Photonic                    & $0.000016555(4)$   & $0.00001139(5)$                \\
 $e$ loop                  & $-0.000144895(1)$     & $-0.000266020(3)$                \\
 $\mu$ loop                  & $-0.00000478850(6)$  & $-0.00000499030(6)$               \\
 $\tau$ loop                & $-0.000000201365(2)$   & $-0.000000196580(2)$             \\
\bottomrule
\end{tabular}}
\caption{Corrections to the decay rate for $\tau\to\ell\nu_\tau\bar\nu_\ell$ in the on-shell scheme, normalized to $\Gamma_0^\tau$, at leading order (LO), next-to-leading order (NLO), and next-to-next-to-leading order (NNLO), the latter excluding HVP. We also provide the ratios of the NLO/LO and NNLO/NLO corrections as well as the breakdown into photonic and the various lepton-loop contributions.}
\label{tab:onshell}
\end{table}

\begin{table}[t]
\scalebox{0.95}{
\renewcommand{\arraystretch}{1.3}
\centering
\begin{tabular}{lcc}\toprule
& $\ell=\mu$ & $\ell=e$\\\midrule
$\tau$ loop ($m_\mu=0$) & \multicolumn{2}{c}{$-0.0364$} \\
$\tau$ loop & $-0.0373$ & $-0.0364$\\
$\mu$ loop ($m_\mu=0$) & \multicolumn{2}{c}{$3.220$}\\
$\mu$ loop & $2.356$ & --\\
$e$ loop ($m_\mu=0$) & \multicolumn{2}{c}{$3.220$}\\
$e$ loop & $2.872$ & $3.220$\\
Photonic ($m_\mu=0$) & \multicolumn{2}{c}{$3.559$}\\
Photonic & $3.087$ & $3.559$ \\\midrule
$\tau$ loop & $-0.0373$ & $-0.0364$\\
$\mu$ loop (on-shell) & $-0.888$ & $-0.925$\\
$\mu$ loop ($\alpha$) & $3.139(7)$ & $3.404(6)$\\
$\mu$ loop ($\mu^+\mu^-$) & $0.104$ & $0.130$\\
$\mu$ loop (total) & $2.355(7)$ & $2.609(6)$\\
$e$ loop (on-shell) & $-26.855$ & $-49.304(1)$\\
$e$ loop ($\alpha$) & $9.068(20)$ & $9.834(17)$\\
$e$ loop ($e^+e^-$) & $20.653$ & $42.681(1)$\\
$e$ loop (total) & $2.866(20)$ & $3.211(17)$\\
Photonic (on-shell) & $3.068(1)$ & $2.111(9)$\\
Photonic ($\ell^+\ell^-$) & $0.019$ & $1.436(1)$\\
Photonic (total) & $3.087(1)$ & $3.547(9)$\\
\bottomrule
\end{tabular}}
\caption{Comparison of two-loop corrections $H_2^{\tau\ell}$ in the $\overline{\text{MS}}$ scheme from analytic (first panel, for $m_\mu=0$ and including mass corrections, where available) and numerical (second panel) calculations. The numerical results are obtained in the on-shell scheme, and the decomposition reflects the reshuffling of terms required to convert to $\overline{\text{MS}}$, by adding the terms from the definition of $\alpha$ and the $\ell^+\ell^-$ final states as detailed in the main text. Numerical errors are only displayed if relevant for the number of digits shown.}
\label{tab:MSbar_onshell}
\end{table}

Alternatively, the calculation can be performed in the on-shell scheme~\cite{Engel:2019nfw}, see Table~\ref{tab:onshell} for the explicit results from {\sc McMule}, in which case the bookkeeping of several contributions changes, as summarized in Table~\ref{tab:MSbar_onshell}. While the $\tau$ loop is the same in both schemes, one needs to add the contribution from the respective logarithm in the running of $\alpha$ as well as the $\ell^+\ell^-$ final states~\cite{vanRitbergen:1999fi}. For the latter, the prescription from Ref.~\cite{vanRitbergen:1999fi} only applies if the $\ell^+\ell^-$ final state has a lepton flavor different from that present in the $\tau\to\ell\nu_\tau\bar\nu_\ell$ decay itself, because otherwise a second Wick contraction arises in the calculation of the decay width that only involves a single Dirac trace. 
This additional single-trace contraction, as can be inferred from the results given in Table~\ref{tab:onshell_ellell} below, needs to be booked as part of the photonic contribution, giving rise to the penultimate row in Table~\ref{tab:MSbar_onshell}. For the electron loop, one observes a significant cancellation among the three contributions, with a sum that reproduces the $\overline{\text{MS}}$ result from the first panel. For the muon loop, final results also agree in all cases that can be compared. The missing mass correction for the electron final state evaluates to $-0.61(1)$, and thus happens to be close to the average of the mass corrections for electron and muon loop in the case of the muon final state.

\section{Hadronic contributions}
\label{sec:hadronic}

The HVP coefficient $H_2^{\tau \ell,\, \text{HVP}}$ can be calculated from the hadronic $R$-ratio $R_\text{had}(s)$ using the kernel function derived in Ref.~\cite{vanRitbergen:1998hn}, see \ref{app:kernel}, where we also spell out the required analytic continuation for the integration range in which the invariant mass squared $s$ of the spectral function is smaller than $4m_\ell^2$. Hadronic cross-section data have been compiled in great detail for the evaluation of the HVP contribution to the anomalous magnetic moment of the muon, see, e.g., Refs.~\cite{Aoyama:2020ynm,Aliberti:2025beg,Aliberti:2024fpq,Hertzog:2025ssc,Davier:2017zfy,Keshavarzi:2018mgv,Colangelo:2018mtw,Hoferichter:2019mqg,Davier:2019can,Keshavarzi:2019abf}. To estimate the uncertainty, we compare the spectral functions from Refs.~\cite{Keshavarzi:2018mgv,Ignatov:VP}. In addition, the dominant uncertainty arises from the tension in the $e^+e^-\to\pi^+\pi^-$ channel between the CMD-3 measurement~\cite{CMD-3:2023alj,CMD-3:2023rfe}  and earlier data sets~\cite{Achasov:2006vp,CMD-2:2006gxt,BaBar:2012bdw,BESIII:2015equ,KLOE-2:2017fda}. We proceed as in Ref.~\cite{Aliberti:2025beg} for the higher-order HVP contribution and assign half the difference between the global fit and the CMD-3-based evaluation (for both using the methodology of Refs.~\cite{Colangelo:2018mtw,Colangelo:2022prz,Stoffer:2023gba}) as an additional source of uncertainty, defining the central value by the mean of both variants. In this way, we first obtain an update of the HVP coefficient for muon decay
\begin{equation}
\label{HVP_muon}
H_2^{\mu e,\, \text{HVP}}=-0.0429(2)(5)[6],
\end{equation}
where the first error refers to the difference between Refs.~\cite{Keshavarzi:2018mgv,Ignatov:VP}, the second one quantifies the tension in the $\pi\pi$ channel, and the third one gives the quadratic sum. The result is consistent with the evaluations from Refs.~\cite{vanRitbergen:1998hn,Eberhart:2026klz}.
The same procedure for the leptonic $\tau$ decay yields
\begin{equation}
H_2^{\tau e,\, \text{HVP}}=-1.189(7)(12)[14],
\end{equation}
while muon mass corrections shift the central value to $-1.191$~\cite{Engel:2019nfw}.

\begin{table}[t]
\scalebox{0.95}{
\renewcommand{\arraystretch}{1.3}
\centering
\begin{tabular}{lcc}\toprule
& $\tau\to \mu \nu_\tau\bar\nu_\mu$ & $\tau\to e\nu_\tau\bar\nu_e$\\\midrule
$\Gamma_0^\tau\  [10^{-13}\GeV]$ & \multicolumn{2}{c}{$4.0484(10)$}\\
$\delta_{\tau\ell}-1\ [10^{-4}]$ & $2.96$ & $2.93$\\ 
$F(\rho_{\tau\ell})$ & $0.9725621(29)$ & $0.999999$\\
$H_1^{\tau \ell}$ & $-1.66834(1)$ & $-1.80979$\\
$H_2^{\tau\ell}-H_2^{\tau\ell,\, \text{HVP}}$ & $8.28$ & $9.35(1)$\\
$H_2^{\tau\ell,\, \text{HVP}}$ & $-1.191(7)(12)[14]$ & $-1.189(7)(12)[14]$\\
$H_3^{\tau\ell}$ & \multicolumn{2}{c}{$-40(1)$}\\\midrule
$\Gamma^\tau\  [10^{-13}\GeV]$ & $3.9227(10)$ & $4.0324(10)$\\
$\Br[\tau\to\ell\nu_\tau\bar\nu_\ell]$ & $0.17301(30)$ & $0.17785(31)$\\
\bottomrule
\end{tabular}}
\caption{SM predictions for the decay widths of $\tau\to \ell\nu_\tau\bar\nu_\ell$. We use particle masses from Refs.~\cite{Tiesinga:2021myr,ParticleDataGroup:2026mpi}: $m_\tau=1.77693(9)\GeV$~\cite{BESIII:2014srs,Anashin:2023sch,Belle-II:2023izd}, $m_\tau/m_\mu=16.8177(9)$, $m_\tau/m_e=3477.37(18)$. For the conversion to the branching ratio, the lifetime $\tau_\tau=290.3(5)\times 10^{-15}\,\text{s}$~\cite{Belle:2013teo,ParticleDataGroup:2026mpi} is used.}
\label{tab:SM_prediction}
\end{table}

\section{Standard-Model prediction for $\tau\to\ell\nu_\tau\bar\nu_\ell$}
\label{sec:SM_prediction}

Our final values for the expansion coefficients in the $\overline{\text{MS}}$ scheme are 
summarized in Table~\ref{tab:SM_prediction}. By far the dominant uncertainty arises from the $\tau$ mass and, for the branching fraction, from its lifetime, both of which are set to improve at Belle II~\cite{Belle-II:2018jsg}, BESIII~\cite{BESIII:2020nme}, a potential Super Tau-Charm Factory~\cite{Achasov:2023gey}, and FCC-ee itself. The dominant uncertainties drop out in the ratio of branching fractions
\begin{equation}
    \frac{\Br[\tau\to \mu \nu_\tau\bar \nu_\mu]}{ \Br[\tau\to e \nu_\tau\bar \nu_e]}
  =0.9727788(29),
\end{equation}
where the error indicates the residual sensitivity to the uncertainty in the $\tau$ mass. 
While our central results are given in Table~\ref{tab:SM_prediction}, 
the detailed breakdown in Table~\ref{tab:MSbar_onshell} allows for easy changes in convention, e.g., if the $\mu^+\mu^-$ final state should not be included in a fully inclusive measurement. In such a case, the muon loop should be treated in the on-shell scheme, changing the two-loop coefficients to
$H_2^{\tau \mu}=5.02$, 
$H_2^{\tau e}=5.82$, respectively, and the value of the fine-structure constant to
$\hat\alpha(m_\tau^2)\to 1/135.301$. 

\section{Radiative and rare decays}
\label{sec:radiative_rare}

We also should mention the related processes $\tau\to\ell\nu_\tau\bar\nu_\ell
\gamma$ and $\tau\to\ell\nu_\tau\bar\nu_\ell \ell'^+\ell'^-$ that of course
contribute to $H_1$ and $H_2$, but can also be measured on their own. In the
case of the former, a 3.5$\sigma$ discrepancy was noted between the NLO
calculation~\cite{Fael:2015gua} for $\ell=e$ (that contributes to $H_2^{\tau
e}$) and a measurement by BaBar~\cite{BaBar:2015ard}. It was later
suggested~\cite{Pruna:2017upz} that this could be due to missing radiative
corrections in the experimental analysis. To briefly summarize the argument made
in Ref.~\cite{Pruna:2017upz}, the $\tau$ radiative branching ratio with photon
energy $E_\gamma \ge 10$\,MeV in the $\tau$ rest frame is computed as
\begin{align}
    \Br = \frac{1}{\epsilon}\frac{N_{\rm obs} - N_{\rm bkg}}{N_{\rm tot}},
\end{align}
where $N_{\rm obs}$ is the number of events surviving the experimental selection
for the decay mode under study, $N_{\rm bkg}$ is the estimated
background, $N_{\rm tot}$ is the total number of decays and $\epsilon$ is the
experimental efficiency. The experimental efficiency has been calculated with a
Monte Carlo simulation based on a LO theory prediction. Using Monte Carlo
simulations equipped with the LO and NLO theory predictions from {\sc McMule}, we have computed
the two efficiencies $\epsilon_{\rm LO}$ and  $\epsilon_{\rm NLO}$, to estimate
the  correction factor $\epsilon_{\rm LO} / \epsilon_{\rm NLO}$ to the
measurement if the more accurate NLO prediction had been used. This amounts to
$0.923(1)$ for $\ell=e$ and $0.989(1)$ for $\ell=\mu$~\cite{Pruna:2017upz},
which would make both experimental measurements consistent with the NLO theory
predictions. The simulation models the main requirements of the BaBar
selection~\cite{Pruna:2017upz} and does not fully account for experimental
details such as trigger efficiency, kinematics-dependent particle identification
performance, photon reconstruction, and additional kinematic requirements.
While repeating the analysis with the NLO simulations would be required
to precisely update the experimental measurements, we expect only minor
adjustments to the estimated correction factors.

\begin{table}[t]
\scalebox{0.95}{
\renewcommand{\arraystretch}{1.3}
\centering
\begin{tabular}{lrr}
\toprule
                 &$ \Gamma[\tau\to \mu\nu_\tau\bar\nu_\mu + \ell^+\ell^-]$ & $\Gamma[\tau\to e\nu_\tau\bar\nu_e + \ell^+\ell^-]$  \\
\midrule
 $\ell = e   $                          & $0.000111435(1)$ & $0.000238034(7)$                         \\
 $\ell = \mu   $                     & $0.000000663285(8)$   & $0.000000701136(7)$                    \\
\midrule
 $\ell = e'/\mu'$                        & $0.000000560102(4)$   & $0.000230287(3)$                   \\
\bottomrule
\end{tabular}}
\caption{Decay rates for the decays with an additional $\ell^+\ell^-$ pair in the final state, normalized to $\Gamma_0^\tau$. The last row gives the putative case of a distinct $\mu'$ or $e'$, respectively, to avoid the contribution from a single Dirac trace.}
\label{tab:onshell_ellell}
\end{table}

\begin{figure}[t]
    \centering
    \includegraphics[trim= 5 5 40 37, clip, width=\linewidth]{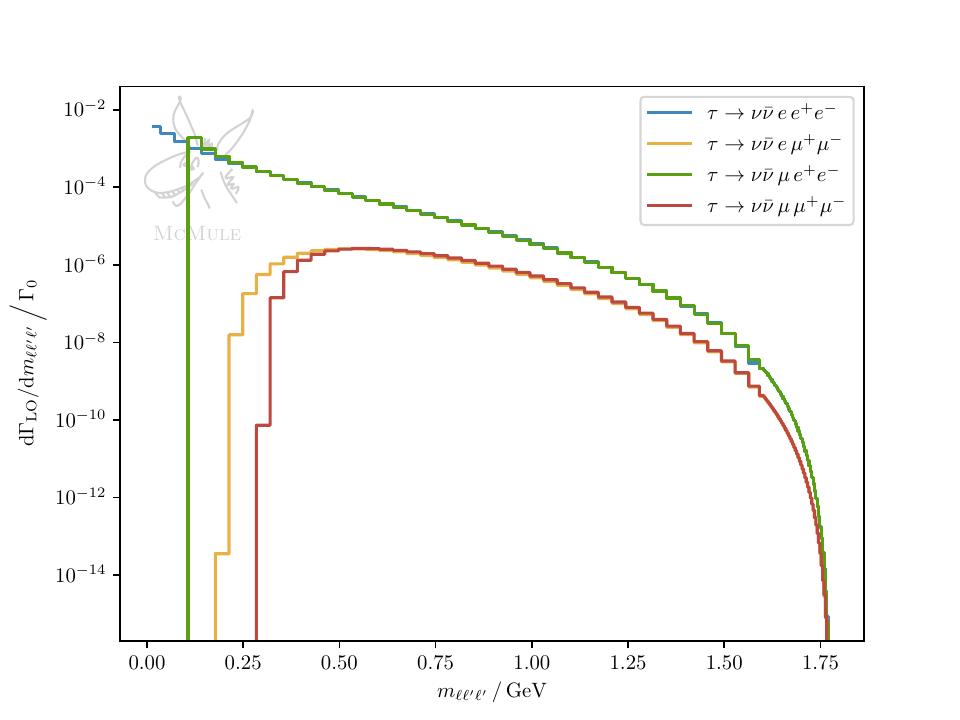}
    \caption{LO decay rate for the process $\tau\to\ell\nu_\tau\bar\nu_\ell \ell'^+\ell'^-$ as a function of the invariant mass $m_{\ell\ell'\ell'}$.}
    \label{fig:m123}
\end{figure}

Next, we turn our attention to the rare decays $\tau\to\ell\nu_\tau\bar\nu_\ell \ell'^+\ell'^-$.
These are related to the process $\mu\to e \nu_\mu\bar\nu_e e^+e^-$ with the important difference that $\ell$ and $\ell'$ no longer need to be the same flavor.
$\mu\to e \nu_\mu\bar\nu_e e^+e^-$ was first considered at NLO in Refs.~\cite{Fael:2016yle,Pruna:2016spf}, while the case of the $\tau$ decay was considered at LO with beyond-SM effects in Ref.~\cite{Flores-Tlalpa:2015xma} and at NLO in Refs.~\cite{Fael:2018vjs,Fael:2019usp}.
For the present discussion LO accuracy is sufficient and we show the branching ratios for all four combinations in Table~\ref{tab:onshell_ellell}.
We also provide results for distinct leptons, and the difference between the results for $\ell=\mu,\mu'$ and $\ell=e,e'$, respectively, allows one to infer the contribution from a single Dirac trace, as required for the conversion to $\overline{\text{MS}}$, see Sec.~\ref{sec:leptonic}.
For completeness and due to its importance for searches for lepton flavor violation, we also provide a differential distribution of this process with respect to the invariant mass of the charged leptons
\begin{align}
    m_{\ell\ell'\ell'} = \sqrt{\big(p_\ell+p_{\ell'^+}+p_{\ell'^-}\big)^2},
\end{align}
for which Fig.~\ref{fig:m123} shows the LO results. 
Especially in the tail where $m_{\ell\ell'\ell'}\simeq m_\tau$, radiative corrections are known to be large, and, if required by the experimental precision, the NLO corrections could be evaluated with the methods of Refs.~\cite{Fael:2018vjs,Fael:2019usp}.

\section{Conclusions}
\label{sec:conclusions}

In this Letter we provided the complete SM predictions for $\tau\to\ell\nu_\tau\bar\nu_\ell$ at NNLO, see Table~\ref{tab:SM_prediction} for our central results. We performed numerical evaluations in the on-shell scheme (including an evaluation of the hadronic contributions), compared to analytic $\overline{\text{MS}}$ results where available, and converted the final results to $\overline{\text{MS}}$ to be able to use known results for the three-loop coefficient. The dominant uncertainties arise from $\tau$ mass and lifetime, but the expansion coefficients are given in a form that allows for easy updates once the precision in the $\tau$ parameters improves. Discarding those uncertainties, the precision is set by the determination of $G_F$ from muon decay, at the level of 1\,ppm, while the remaining uncertainty in the NNLO coefficient and the total size of the estimated third-order contribution amount to 0.1\,ppm and 0.5\,ppm, respectively. 

We also commented on radiative and rare decays, which must be considered in the conversion between $\overline{\text{MS}}$ and on-shell schemes, clarifying the origin of an apparent tension in the radiative channel and providing differential distributions in the three-charged-lepton invariant mass as required for background estimates in the search for lepton-flavor-violating $\tau$ decays.

\section*{Data availability}

Some of the numerical data in this paper was generated using the {\sc McMule} Monte Carlo code~\cite{Banerjee:2020rww} and the raw data is available at \cite{McMuledata}.

\section*{Acknowledgments}

We thank Peter Stoffer for discussions on LEFT anomalous dimensions and Aidan Wright for an independent cross check of the hadronic contributions.  
Financial support by the SNSF (Project No.\ TMCG-2\_213690) is gratefully acknowledged. This research was supported by the Munich Institute for Astro-, Particle and BioPhysics (MIAPbP) which is funded by the Deutsche Forschungsgemeinschaft (DFG, German Research Foundation) under Germany's Excellence Strategy -- EXC-2094 -- 390783311.

\appendix

\section{Kernel function for hadronic contributions}
\label{app:kernel}

The contribution from internal hadronic (had) or leptonic ($\ell'$) loops at $\Order(\alpha^2)$ can be expressed in the form~\cite{vanRitbergen:1998hn} 
\begin{equation}
  \Gamma_R^{\ell}=\Gamma_0^{\ell}\bigg(\frac{\alpha}{\pi}\bigg)^2\int_{z_\text{thr}}^\infty dz\, K(z) R\big(m_\ell^2 z\big),
\end{equation}
where the integrand is determined by the respective $R$-ratio 
\begin{align}
R_\text{had}(s)&=\frac{3s}{4\pi\alpha^2}\sigma\big[e^+e^-\to\text{hadrons}\big]=-\frac{3}{\alpha}\Im \Pi_\text{had}(s),\notag\\
R_{\ell'}(s)&=\bigg(1+\frac{2m_{\ell'}^2}{s}\bigg)\sqrt{1-\frac{4m_{\ell'}^2}{s}},
\end{align}
the thresholds are $z_\text{had}=M_{\pi^0}^2/m_\ell^2$, $z_{\ell'}=4m_{\ell'}^2/m_\ell^2$, and the kernel function reads 
\begin{align}
    3z\, K(z)&=-\frac{1}{432}\big(513+2540z-180z^2-144z^3\big)\\
    &-\frac{1}{72}\big(120+172z-45z^2-12z^3\big)\log z\notag\\
    &+\frac{1}{24\sigma}\big(132-62z-11z^2+4z^3\big)\log\frac{1+\sigma}{1-\sigma}\notag\\
    &-\frac{1}{6}\big(6+16z-18z^2+z^4\big)\bigg(\frac{\pi^2}{6}+\log\frac{1-\sigma}{2}\log\frac{1+\sigma}{2}\bigg)\notag\\
    &+\frac{z^2}{6}\big(12-2z-z^2\big)\sigma\bigg[\text{Li}_2\bigg(\frac{1-\sigma}{2}\bigg)-\text{Li}_2\bigg(\frac{1+\sigma}{2}\bigg)\bigg],\notag
\end{align}
where $\sigma=\sqrt{1-4/z}$. In particular, in the case of $\ell'=\ell$, this representation reproduces the analytic result~\cite{vanRitbergen:1998hn} 
\begin{equation}
  \Gamma_\ell^{\ell}=\Gamma_0^\ell \bigg(\frac{\alpha}{\pi}\bigg)^2\bigg(\frac{16987}{576} - \frac{85}{216}\pi^2  - \frac{64}{3}\zeta(3)\bigg),
\end{equation}
while in the other cases it yields an efficient tool for numerical evaluations. For applications to light internal lepton loops and the hadronic contribution the kernel function needs to be analytically continued in part of the integration region, resulting in 
\begin{align}
 3z\,\tilde K(z)&=-\frac{1}{432}\big(513+2540z-180z^2-144z^3\big)\\
    &-\frac{1}{72}\big(120+172z-45z^2-12z^3\big)\log z\notag\\
    &+\frac{1}{12\tilde \sigma}\big(132-62z-11z^2+4z^3\big)\arctan\tilde\sigma\notag\\
    &-\frac{1}{6}\big(6+16z-18z^2+z^4\big)\bigg(\frac{\pi^2}{6}+\frac{1}{4}\log^2 z+\arctan^2\tilde\sigma\bigg)\notag\\
    &+\frac{z^2}{3}\big(12-2z-z^2\big)\tilde\sigma\notag\\
    &\qquad\times\bigg[
    \frac{\log z}{2}\bigg(\arctan\tilde\sigma -\pi\bigg)+\Im \text{Li}_2\bigg(\sqrt{z}\,e^{i\arctan\tilde\sigma}\bigg)\bigg],\notag
\end{align}
where $\tilde\sigma=\sqrt{4/z-1}$.

\bibliographystyle{apsrev4-1_mod}
\balance
\biboptions{sort&compress}
\bibliography{tau}

\end{document}